\documentclass[twocolumn,prl,aps,showpacs,preprintnumbers,superscriptaddress]{revtex4}

\usepackage{graphicx}
\usepackage{amsmath}

\newcommand{\be}{\begin{equation}}
\newcommand{\ee}{\end{equation}}
\newcommand{\bea}{\begin{eqnarray}}
\newcommand{\eea}{\end{eqnarray}}
\newcommand{\HH}{{\cal H}}

\newcommand{\la}{\langle}
\newcommand{\ra}{\rangle}

\newcommand{\lp}{\left(}
\newcommand{\rp}{\right)}

\renewcommand{\epsilon}{\varepsilon}

\begin{document}
\title{Quantum Phase Tomography of a Strongly Driven Qubit}

\author{M.~S.~Rudner}
\affiliation{Department of Physics, Massachusetts Institute of Technology, Cambridge MA 02139}
\author{A.~V.~Shytov}
\affiliation{Department of Physics,
University of Utah,
Salt Lake City, UT 84112}
\author{L.~S.~Levitov}
\affiliation{Department of Physics, Massachusetts Institute of Technology, Cambridge MA 02139}
\affiliation{Kavli Institute for Theoretical Physics, University of California, Santa Barbara, CA 93106}
  \author{D.~M.~Berns}
  \affiliation{Department of Physics, Massachusetts Institute of Technology, Cambridge MA 02139}
 \author{W.~D.~Oliver}
  \affiliation{MIT Lincoln Laboratory, 244 Wood Street, Lexington, MA 02420}
 \author{S.~O.~Valenzuela}
  \affiliation{MIT Francis Bitter Magnet Laboratory, Cambridge, MA 02139}
 \author{T.~P.~Orlando}
  \affiliation{Department of Electrical Engineering and Computer Science, Massachusetts Institute of Technology, Cambridge, MA 02139}


\begin{abstract}
The interference between repeated Landau-Zener transitions in a qubit swept through an avoided level crossing results in St\"uckelberg oscillations in qubit magnetization. The resulting oscillatory patterns are a hallmark of the coherent strongly-driven regime in qubits, quantum dots and other two-level systems. The two-dimensional Fourier transforms of these patterns are found to exhibit a family of one-dimensional curves in Fourier space, in agreement with recent observations in a superconducting qubit. We interpret these images
in terms of time evolution of the quantum phase of qubit state
and show that they can be used to probe dephasing mechanisms in the qubit.
\end{abstract}
\maketitle
 \vspace{-10mm}

Superconducting Josephson devices
recently emerged as a platform for exploring 
coherent quantum dynamics 
in solid state systems~\cite{Mooij05}. 
Due to their macroscopic dimensions, these devices feature strong coupling to RF fields~\cite{Chiorescu04,Wallraff04,Schuster07}, 
and can be used to study new
quantum phenomena associated with strong driving 
such as 
Rabi oscillations in the multiphoton
regime~\cite{Nakamura01,Saito06a}, 
Landau-Zener-St\"uckelberg-type (LZS)
interference~\cite{Oliver05,Sillanpaa06,Berns06}, Bloch oscillations~\cite{Bloch_oscillations}, and 
qubit-photon dressed states~\cite{Wilson07}.


\begin{figure}[h]
\includegraphics[width=3.2in]{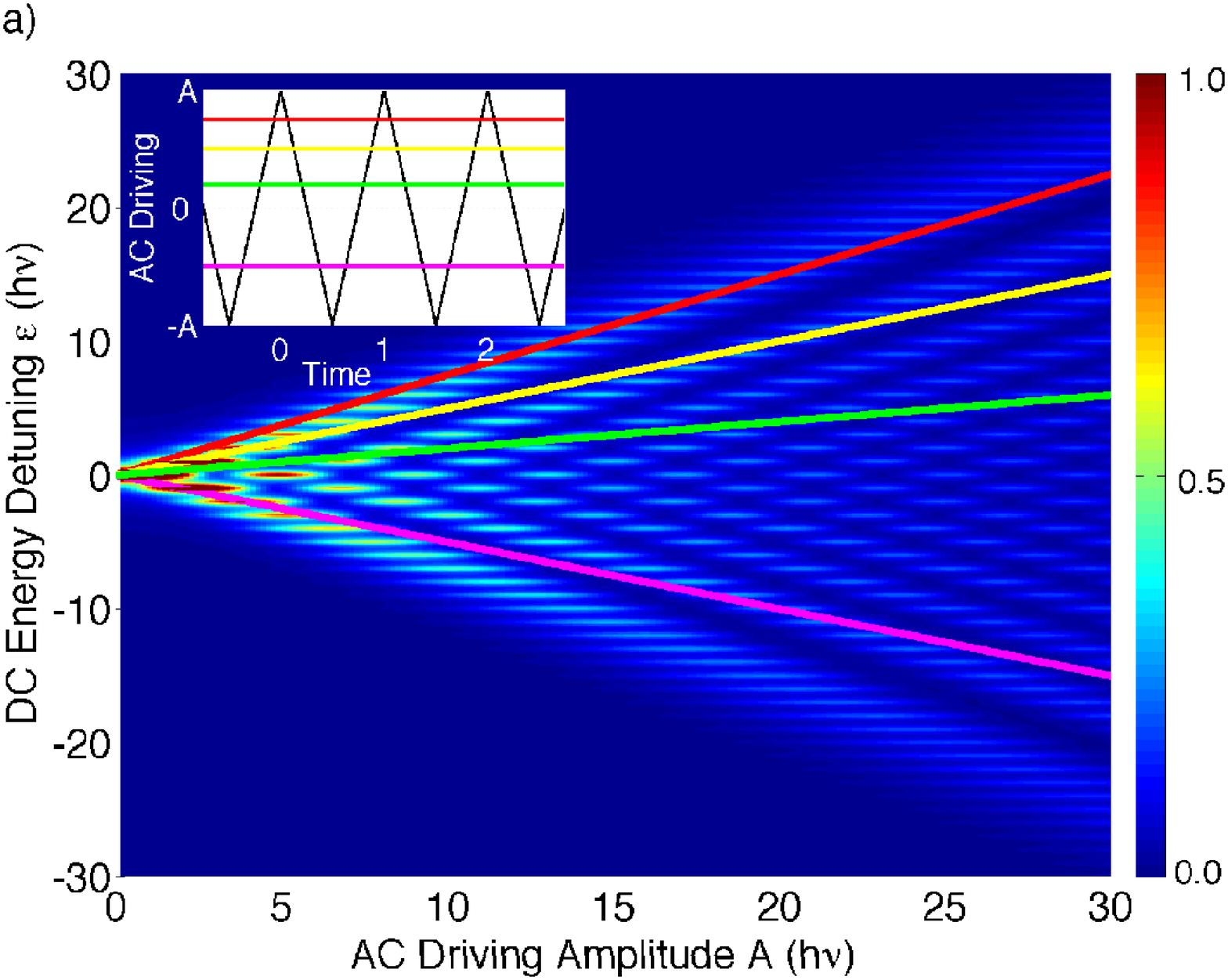}
\includegraphics[width=3.2in]{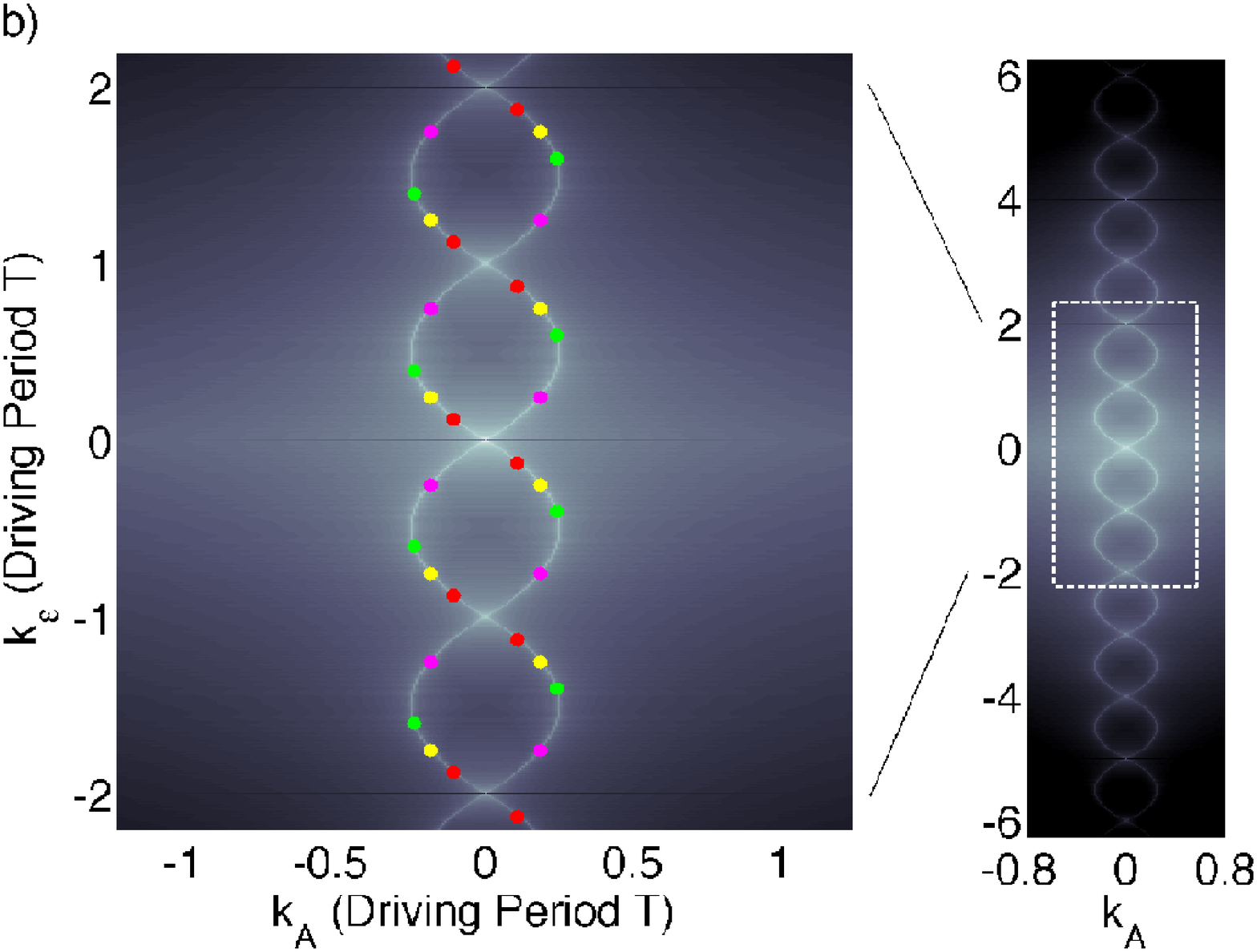}
\caption[t]{Tomographic imaging of qubit phase evolution. 
The pattern of LZS oscillations in (a) the transition rate (\ref{eq:W}) 
and (b) its Fourier transform, which exhibits a family of parabolic arcs (\ref{eq:parabolas}) forming lemon-shaped ovals. 
Along each arc, $k_\epsilon$ and $k_A$ represent the time separation $t_2 - t_1$ and phase gain $\phi^g_{12}$, Eq.(\ref{eq:(k_epsilon, k_A)-2}), between subsequent level crossings, respectively.
The Fourier intensity at each such point 
is mapped from the region near a ray $\epsilon/A=u$ in the $(A,\epsilon)$ plane,
where the parameters $A$, $\epsilon$ all yield the same time interval and phase gain between level crossings. 
Four of these rays and the corresponding points in Fourier space are shown in matching colors.
A Sawtooth-like driving signal (inset) was used, with the decoherence rate $\Gamma_2=\frac14\omega$ in (\ref{eq:W}).
}
 \label{fig1}
\vspace{-5mm}
\end{figure}


In the LZS regime~\cite{Oliver05,Sillanpaa06,Berns06}, the qubit undergoes repeated Landau-Zener (LZ) transitions at a level crossing,
with adiabatic evolution between crossings~\cite{Izmalkov04}.
Interference between subsequent LZ transitions results in an oscillatory dependence of qubit magnetization in the final state on the detuning from the level crossing and the driving amplitude~\cite{Oliver05,Ashhab07}. 
The LZS effect is related to  
earlier observations of photon-assisted
transport in quantum dots~\cite{Kouwenhoven94,Naber06}
and in superconducting systems~\cite{Tien63a,NakamuraTsai99},
which exhibit 
multiphoton sidebands with oscillatory dependence on RF field amplitude.
Although the observed oscillations washed out more quickly at high RF power in those devices than in the qubits \cite{Oliver05,Sillanpaa06,Berns06}, in all cases the oscillations 
originated from the LZS interference effect. 

A new regime of strong driving  
was reported in a recent work~\cite{amplitude_spectroscopy}, 
in which a qubit was driven through a manifold of several states 
spanning a wide energy range up to 120\,GHz. 
The observed LZS interference indicated that even such strong driving was not detrimental for coherence. 
The change in qubit magnetization induced by the driving pulse exhibited complex checkerboard-like patterns in the two-dimensional phase space parameterized by the DC magnetic flux and RF driving amplitude. 
These patterns displayed a multiscale character, with multiphoton resonance lines on the finest 
scale and LZS interference fringes on a larger scale, 
and additional complexity due to resonance and interference effects involving several pairs of energy levels during each pulse.

In an attempt to better understand the observed patterns, the authors of Ref.\,\cite{amplitude_spectroscopy} employed 
a two-dimensional Fourier transform (FT). 
Unexpectedly, the FT revealed a highly ordered structure 
of {\it one-dimensional arcs} joined together to form lemon-shaped ovals in Fourier space,
in contrast to the familiar Bragg peaks in the Fourier images of periodic patterns. 
Most surprisingly, these arcs 
were found to connect the high and low wavenumber regions, which are
associated with the multiphoton resonances and LZS interference fringes. 

In this article we explain the lemon-shaped structures observed in~\cite{amplitude_spectroscopy}, first using a quasiclassical phase argument and then fully microscopically. 
We analyze the FT of the transition rate (see Fig.\ref{fig1}), which can be measured using short excitation pulses
~\cite{Berns06}. Then we consider the FT of qubit 
magnetization produced by long pulses. 

Our analysis reveals a relation between the lemon-shaped structures and the coherent dynamics of the qubit. 
In fact, because the Fourier transform inverts the energy variable and maps it onto the time variable (scaled by $\hbar$), we find that the lemon arcs can be interpreted as an image of the time dependence of the quantum phase of the qubit (see Eq.(\ref{eq:phi(epsilon,A)}) below). 
This relation, as we shall discuss, can be exploited to probe fundamental aspects of qubit dynamics such as decoherence and dephasing, and to measure the decoherence times $T_2$ and $T_2^\ast$. 

The intensity
of each point on the curve in Fourier space arises from a particular ray-like section of the LZS pattern (see Fig.\ref{fig1}), with the section direction in one-to-one correspondence with the time interval beween level crossings.
The section-by-section mapping 
to Fourier space is reminiscent of tomographic imaging, and realizes a ``tomogram'' of the time evolution of the qubit phase.

Employing the FT to image quantum phase is 
familiar from the work on the mesoscopic Aharonov-Bohm effect~\cite{Webb85,Marcus92},
which used the dependence of conductance on magnetic field. 
In our approach, however, the {\it time dependence} of the phase is reconstructed using a two-dimensional FT where 
the axis associated with the 
detuning from qubit level crossing plays the role of time.

We also note that in recent work \cite{Katz06,Steffen06a,Steffen06b} a tomographic reconstruction of the Wigner function on the Bloch sphere was performed.
The procedure used in Refs.\cite{Katz06,Steffen06a,Steffen06b}, which employs controlled rotations of the qubit state following its Rabi oscillations, is different from that used in the present work. Our image of qubit time evolution is obtained in 
Fourier space. Also, because of the nature of the LZS effect, it only provides information about the relative phase of the qubit $|0\ra$ and $|1\ra$ states, rather than the entire Bloch vector.

To emphasize aspects common to different experiments that have used harmonic~\cite{Oliver05,Sillanpaa06,Berns06,Wilson07,amplitude_spectroscopy}, sawtooth-like~\cite{Bloch_oscillations}, and bi-harmonic~\cite{Bylander} RF driving, we consider a generic periodic driving of the qubit:
\be \label{Eq:Hamiltonia} 
\HH = -\frac{\hbar}2 
\begin{pmatrix}
h(t) & \Delta \cr \Delta &
-h(t)
\end{pmatrix} 
,\quad
h(t)=\epsilon-A g(t)
,
\ee
where $h(t)$ is the
energy detuning from an avoided crossing, 
periodically modulated by the driving
field $g(t)=g(t+T)$ with amplitude $A$ and 
zero mean, $\oint g(t)dt=0$. 
For simplicity here we focus on the case when $g(t)$ has one maximum and one minimum per period.

Away from the level crossing, the qubit evolves adiabatically as a superposition of the states $|0\ra$ and $|1\ra$. The LZS interference 
can be expressed~\cite{Oliver05} through the relative phase of the states $|0\ra$ and $|1\ra$  gained between subsequent passages through a level crossing: 
\be\label{eq:phi(epsilon,A)}
\phi(A, \epsilon)=\int\limits_{t_1}^{t_2} h(t)dt =\epsilon(t_2-t_1)-A\int\limits_{t_1}^{t_2}g(t)dt
.
\ee
The times $t_{1,2}$ 
of level crossing are the solutions to 
\be\label{eq:Ag(t)=epsilon}
Ag(t)=\epsilon
\qquad
(g_{\rm min}<\epsilon/A<g_{\rm max})
,
\ee
represented graphically in the inset of Fig.\ref{fig1}a by the intersections between 
lines of fixed detuning and 
the driving signal.
Quasiclassically, the LZS 
contrast 
can be modeled by a sum of functions $\cos(\phi(A, \epsilon))$, one per each choice of $t_{1,2}$ in \eqref{eq:phi(epsilon,A)}.
We consider a position-dependent wavevector 
\be\label{eq:(k_epsilon, k_A)-1}
(k_A,k_\epsilon)= \pm (\nabla_A \phi(A, \epsilon), \nabla_\epsilon \phi(A, \epsilon)), 
\ee
where $\pm$ accounts for the contributions of $e^{\pm i \phi(A, \epsilon)}$. 
Evaluating the derivatives in (\ref{eq:(k_epsilon, k_A)-1}) and,
noting that the net contributions of $\nabla_\epsilon t_{1,2}$ and $\nabla_A t_{1,2}$ vanish due to (\ref{eq:Ag(t)=epsilon}), gives
\be
\label{eq:(k_epsilon, k_A)-2}
(k_A,k_\epsilon)= \pm (-\phi^g_{12}, t_2-t_1)
,\quad
\phi^g_{12}=\int_{t_1}^{t_2}g(t)dt
.
\ee
Crucially, Eq.(\ref{eq:(k_epsilon, k_A)-2}) defines
{\it a curve} parameterized by a single variable $u=\epsilon/A$,
which is the only parameter upon which
the 
times $t_{1,2}$ found in Eq.(\ref{eq:Ag(t)=epsilon}) depend.

We illustrate this mapping by an example of sawtooth driving (Fig.\ref{fig1} inset) with $g(t)$ 
linear between the points
\be\label{eq:sawtooth}
g(nT)=-g((n\pm {\textstyle \frac12})T)=1
.
\ee
In Fig.\ref{fig1}, 
the points in Fourier space arising from different 
$\epsilon/A$ sections are denoted by dots with colors matching those of the corresponding rays $\epsilon/A=u$ in the $(A,\epsilon)$ plane and of the constant detuning lines in the inset. 
The $k_\epsilon$ and  $k_A$ coordinates of those points correspond to the time separation (Fig.1a inset) and the phase gain, Eq.(\ref{eq:(k_epsilon, k_A)-2}), between subsequent level crossings. 
In this way, 
the curves (\ref{eq:(k_epsilon, k_A)-2}) reproduce the time evolution of qubit phase.

As shown in Fig.\ref{fig1}, each ray maps to a family of points (\ref{eq:(k_epsilon, k_A)-2}). 
The reason for this multiplicity is that, besides the sign $\pm$ in Eq. (\ref{eq:(k_epsilon, k_A)-2}), Eq.(\ref{eq:Ag(t)=epsilon}) has multiple solutions $t_1'=t_1+n_1T$ and $t_2'=t_2+n_2T$ for each $\epsilon$ and $A$, where $T$ is the period of driving and $n_{1,2}$ are arbitrary integers.
Because 
$\oint g(t) dt = 0$, all such solutions yield the same phase gain $\phi^g_{12}$ and the same value of $k_A$.
However, the corresponding values of $k_\epsilon$ are displaced by $(n_2-n_1)T$, generating the periodic family of arcs shown in Fig.\ref{fig1}b.

\begin{figure}[t]
\includegraphics[width=3.4in]{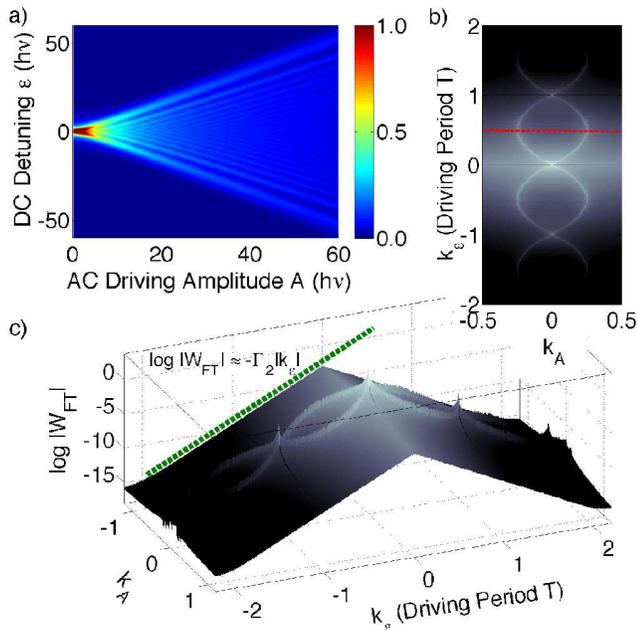}
\caption[t]{Transition rate (\ref{eq:W}) and its FT for sawtooth-like driving for $\Gamma_2=\omega$, four times larger than in Fig.\ref{fig1}. A 3D projection plot of Fourier intensity is shown to illustrate the exponential decay $W_{FT}(k_A,k_\epsilon)\propto e^{-\Gamma_2 k_\epsilon}$.
}
 \label{fig2}
\vspace{-5mm}
\end{figure}

Another 
class of solutions to Eq.(\ref{eq:Ag(t)=epsilon}) 
describes subsequent passages through the level crossing {\it in the same direction}:
$t_2=t_1+nT$. 
In this case $t_1$ is unconstrained and, because 
zero phase is gained over a single driving period, 
we obtain a discrete set of points $(k_A,k_\epsilon)=(0,nT)$ irrespective of 
$\epsilon, A$. 
As displayed most clearly in Fig.\ref{fig2}c, the FT intensity indeed exhibits peaks at $k_\epsilon = nT$.
The peak positions $k_\epsilon=nT$ agree with the spacing $\hbar\omega$ between multiphoton resonances in the $(A,\epsilon)$ plane. 


To find the form of the curves in Fig.\ref{fig1}b, we solve 
Eq.(\ref{eq:Ag(t)=epsilon}) for the case of sawtooth
driving, Eq.(\ref{eq:sawtooth}).
Without loss of generality we select $-T/2<t_1<0<t_2<T/2$ and find 
\be\label{eq:t_1,2}
t_2=-t_1=\tau/2
,\quad 
\tau\equiv T(A-\epsilon)/2A
.
\ee
Evaluating the phase 
$\phi^g_{12}=\int_{t_1}^{t_2}g(t)dt=\frac14(1-\epsilon^2/A^2)T$, 
we obtain parabolic arcs 
in Fourier space:
\be\label{eq:parabolas}
(k_A,k_\epsilon) = \pm \lp -(1-\tau/T)\tau, \tau + nT\rp
,
\ee
$0<\tau<T$, where the term $nT$ was added to $k_\epsilon$ to account for the multiple solutions to Eq.(\ref{eq:Ag(t)=epsilon}) discussed above. 

Similarly, in the case of harmonic driving, the solutions of Eq.(\ref{eq:Ag(t)=epsilon}) are $t_2=-t_1=\frac1{\omega}\arccos(\epsilon/A)$, which gives the phase 
$\phi^g_{12}=\int_{t_1}^{t_2} \cos(\omega t) dt=(2/\omega)\sqrt{1-\epsilon^2/A^2}$. Substituting these results into Eq.(\ref{eq:(k_epsilon, k_A)-2}) we obtain the sinusoids
\be\label{eq:sinusoids}
\omega k_A/2 =\pm
\sin\lp \omega k_\epsilon/2\rp
\ee
which were 
observed in Ref.\,\cite{amplitude_spectroscopy}.


\begin{figure}[t]
\includegraphics[width=3.3in]{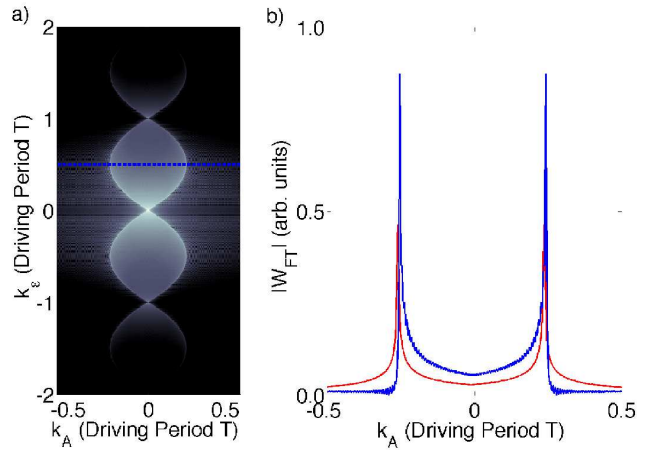}
\caption[t]{Test of the result (\ref{eq:sinusoidal_lemon}) for the Fourier intensity distribution in the lemon. a) FT of the transition rate in Fig.\ref{fig2} after the image was doubled by extending 
$W$ to negative $A$ as $W(-A,\epsilon)=W(A,\epsilon)$.
b) Intensity in a mid-lemon section (blue line) compared with a similar section through a lemon in FT of the undoubled image in Fig.\ref{fig2} (red line).
}
 \label{fig3}
\vspace{-5mm}
\end{figure}


Now we turn to a microscopic analysis of qubit dynamics based on the Hamiltonian (\ref{Eq:Hamiltonia}) to which we add classical noise to model decoherence: $\tilde h(t)=h(t) + \delta\epsilon(t)$. The transitions between qubit states $|0\ra$ and $|1\ra$ can be analyzed most easily in a rotating frame where
\be
\HH =
 -\frac{\hbar}2 
\begin{pmatrix} 
0 & \Delta(t) \cr \Delta^\ast(t) & 0
\end{pmatrix}
,\quad
\Delta(t)=\Delta e^{-i\tilde\phi(t)}
,
\ee
with $\tilde\phi(t)=\int^t_0 \tilde h(t')dt' $. Perturbation theory in $\Delta$  
yields
the rate of transitions between the states $|0\ra$ and $|1\ra$:
\be\label{eq:Wgeneral}
    W =
        \lim_{\delta t \Gamma_2\gg 1} \frac{\Delta^2}{4\delta t} \iint_t^{t+\delta t} \la e^{-i\tilde \phi(t_1)} e^{i\tilde \phi(t_2)}\ra_{\delta \epsilon} dt_1 dt_2
    ,
\ee
where $\Gamma_2=\frac1{T_2}$ is the decoherence rate. We average over $\delta\epsilon(t)$ using the white noise
model: $\la
e^{i\delta\phi(t_2)-i\delta\phi(t_1)}\ra_{\delta \epsilon}=e^{-\Gamma_2|t_1-t_2|}$, 
where $\delta\phi(t)=\int_0^t\delta\epsilon(t')dt'$.

To find the rate $W$ in closed form, we use the Fourier series  $e^{i\phi(t)}=e^{i\epsilon t}\sum_m f_m e^{-im\omega t}$, where the coefficients $f_m$
can be expressed through the error function of complex argument for the case of sawtooth driving, or Bessel functions for the case of harmonic driving~\cite{Berns06}. 
Using the appropriate Fourier series in (\ref{eq:Wgeneral}) and performing the integration over $t_1$ and $t_2$, we obtain the expression
\be\label{eq:W}
    W(\epsilon,A)
    = \frac{\Delta^2}2\sum_{m=-\infty}^\infty \frac{\Gamma_2
    |f_m|^2}{(\epsilon-\omega m)^2+\Gamma_2^2}
.
\ee
At $\omega \gtrsim 2\pi \Gamma_2$
this expression describes non-overlapping resonances (see Fig.\ref{fig1}),
while at $\omega \lesssim 2\pi \Gamma_2$ it describes the partially dephased regime of Ref.\,\cite{Berns06} (see Fig.\ref{fig2}).

To evaluate the Fourier transform of the transition rate $W_{FT}(k_A,k_\epsilon)=\iint_{-\infty}^\infty e^{-iA k_A -i\epsilon k_\epsilon} W(\epsilon,A)d\epsilon dA$, it is convenient to return to expression (\ref{eq:Wgeneral}). 
Because the phase $\phi(t)=\epsilon t-\int_0^t Ag(t')dt'$ is linear in $\epsilon$ as well as in $A$, we can easily bring the Fourier transform of (\ref{eq:Wgeneral}) to the form
\[
a \iint_t^{t+\delta t} 
\delta(k_\epsilon+t_1-t_2)\delta(k_A+ \phi^g_{12}) 
e^{-\Gamma_2|t_1-t_2|} dt_1 dt_2
\]
with $a=\Delta^2(2\pi)^2/4\delta t$ and 
$\phi^g_{12}$ defined in (\ref{eq:(k_epsilon, k_A)-2}).
This result can be simplified by performing the integration over $t_2$ with the help of the delta function $\delta(k_\epsilon+t_1-t_2)$, giving
\be\label{eq:W_FT-2}
W_{FT}(k_A,k_\epsilon)=\textstyle{\pi\over2}\Delta^2\omega
e^{-\Gamma_2|k_\epsilon|}
\oint\delta(k_A+\phi^g_{12}) 
 dt_1 
,
\ee
where $t_2=t_1+k_\epsilon$.
Result (\ref{eq:W_FT-2}) illustrates the effect of dephasing on the lemon structure through 
the prefactor $e^{-\Gamma_2|k_\epsilon|}$ (see Fig.\ref{fig2}c), 
which arises from the exponential decay in time 
$e^{-|t_2-t_1|/T_2}$, and is consistent with 
the interpretation of $k_\epsilon$ as a time variable. 

It is instructive to compare this behavior with the effect of ensemble averaging, modeled by random offsets $\delta\epsilon$ with a gaussian distribution. Because the phase factors in (\ref{eq:Wgeneral}) are linear in $\epsilon$, the ensemble-averaged FT is
\be\label{eq:T2_T2*}
\la W_{FT}(k_A,k_\epsilon)\ra_{\rm ens}\propto e^{-\Gamma_2|k_\epsilon|} e^{-\frac12\lambda k_\epsilon^2}
,\quad
\lambda=\la\delta\epsilon^2\ra
.
\ee
Through this dependence, instrinsic dephasing and ensemble averaging, i.e. $T_2$ and $T_2^\ast$, can be distinguished.  

The lemon boundary obtained from (\ref{eq:W_FT-2}) for a generic $g(t)$ agrees with the quasiclassical result (\ref{eq:(k_epsilon, k_A)-2}). Indeed, the range of $k_A$ for which FT intensity is nonzero, at a fixed $k_\epsilon$, are determined by the extrema of the function $\phi^g_{12}$ in $t_1$. Writing $\delta_{t_1}\phi^g_{12}=g(t_2)-g(t_1)=0$ we recover Eq.(\ref{eq:Ag(t)=epsilon}).


\begin{figure}[t]
\includegraphics[width=3.4in]{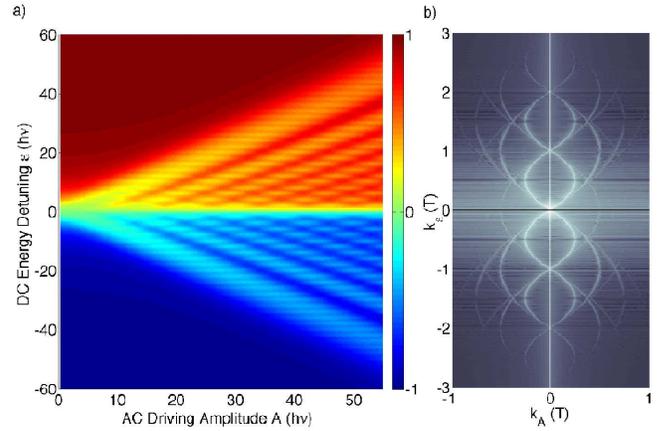}
\caption[t]{Qubit magnetization and its FT. 
Shown is the magnetization of a qubit driven to saturation:
$m=(\Gamma_1-\Gamma'_1)/(2W+\Gamma_1+\Gamma'_1)$,
$\Gamma'_1=\Gamma_1 e^{-\epsilon/k_{\rm B}T}$,
where $\Gamma_1$ ($\Gamma'_1$) is the down (up) relaxation rate~\cite{Berns06}.
Results are shown for sawtooth driving with parameter values: decoherence rate $\Gamma_2=\frac12\omega$, temperature $k_{\bf B}T=1.2\hbar\omega$, relaxation rate $\Gamma_1=8\cdot 10^{-5} \omega$, frequency $\nu=400\,{\rm MHz}$, level splitting $\Delta=12\,{\rm MHz}$.}
 \label{fig4}
\vspace{-5mm}
\end{figure}


For the case of harmonic driving, $g(t)=\cos\omega t$, we can evaluate (\ref{eq:W_FT-2}) by noting that 
$\phi^g_{12}=(\sin(\omega t_2)-\sin(\omega t_1))/\omega=(2/\omega)\sin(\frac12 \omega k_\epsilon)\cos (\omega(t_1+\frac12 k_\epsilon))$. The integral over $t_1$ in (\ref{eq:W_FT-2}) then yields 
\be\label{eq:sinusoidal_lemon}
W_{FT}(k_A,k_\epsilon)=\frac{\Delta^2\omega\, e^{-\Gamma_2|k_\epsilon|}}{2 \sqrt{
\frac4{\omega^2}\sin^2(\frac12 \omega k_\epsilon)-k_A^2}}
\ee
for $|k_A|<\frac2{\omega}|\sin(\frac12 \omega k_\epsilon)|$, and zero elsewhere. We see that $W_{FT}(k_A,k_\epsilon)$ is concentrated inside the region bounded by the sinusoids (\ref{eq:sinusoids}) with square root singularities at the boundary. 
Similar behavior with a square root singularity in FT intensity is obtained for the sawtooth case, as illustrated in Fig.\ref{fig3}. 
Because Eq.(\ref{eq:sinusoidal_lemon}) is derived with the FT taken over $-\infty < A < \infty$, 
the LZS pattern in Fig.\ref{fig2} had to be doubled to obtain the FT in Fig.\ref{fig3}.

Finally, lemon structures are also exhibited by the FT of the qubit steady-state population.
The lemon arcs with multiple periods, clearly visible in Fig.\ref{fig4}, arise because of a nonlinear dependence of saturated population on $W$, with quadratic nonlinearity giving double period, cubic nonlinearity giving triple period, etc. This multiplicity of periods was also observed in the data~\cite{amplitude_spectroscopy}. 


In conclusion, FT-based tomography of two-dimensional LZS patterns is a general technique that offers a way to image the quantum phase evolution of qubits and other quantum systems.
In the simplest case of a driving signal with just one maximum and one minimum per period, 
we predict a chain-like lemon structure in Fourier space which is in perfect agreement with recent 
observations.

We acknowledge partial support from W. M. Keck
Foundation Center for Extreme Quantum Information
Theory. The work of M. R. was supported by DOE
CSGF, Grant No. DE-FG02-97ER25308. The work at Lincoln Laboratory was sponsored by the US DoD under Air Force Contract No. FA8721-05-C-0002. The views and conclusions contained in this document are those of
the authors and should not be interpreted as representing the official policies, either expressly or implied, of the U.S. Government.

\end{document}